\documentclass{mn2e}

\newcommand{\hi}{H\,{\sc i}}

\newcommand{\ho}{$H_{0}$}
\newcommand{\etal}{et al.}

\newcommand{\ltsim}{\raisebox{-1mm}{$\stackrel{<}{\sim}$}}

\newcommand{\gtsim}{\raisebox{-1mm}{$\stackrel{>}{\sim}$}}

\newcommand{\mdot}{M$_{\odot}$}
\newcommand{\nh}{$N_{\rm H}$}

\newcommand{\dgr}{$^{\circ}$}

\newcommand{\eg}{e.g.}
\newcommand{\ie}{i.e.}

\def\arcm{\hbox{$^\prime$}}
\def\arcs{\arcm\hskip -0.1em\arcm}

\title[{\em Chandra} Observations of the {\em Mice}]{{\em Chandra}
Observations of the {\em Mice}}

\author[A.M. Read]{Andrew
M. Read$^{1}$ \\  
$^{1}$ School of Physics and Astronomy, University of Birmingham, 
Edgbaston, Birmingham, B15 2TT, UK\\
(E-mail: amr@star.sr.bham.ac.uk)} 

\date{Accepted ..............................; 
Received ..............................; 
in original form ..............................}

\begin{document}

\maketitle

\begin{abstract} 

Presented here are high spatial and spectral resolution Chandra X-ray
observations of the famous interacting galaxy pair, the Mice, a system
similar to, though less evolved than, the well-known Antennae
galaxies. Previously unpublished ROSAT HRI data of the system are also
presented.

Starburst-driven galactic winds outflowing along the minor axis of
both galaxies (but particularly the northern) are observed, and
spectral and spatial properties, and energetics are presented. That
such a phenomenon can occur in such a rapidly-evolving and turbulent
system is surprising, and this is the first time that the very
beginning $-$ the onset, of starburst-driven hot gaseous outflow in a
full-blown disk-disk merger has been seen.

Point source emission is seen at the galaxy nuclei, and within the
interaction-induced tidal tails. Further point source emission is
associated with the galactic bar in the southern system. A comparison
of the source X-ray luminosity function and of the diffuse emission
properties is made with the Antennae and other galaxies, and evidence
of a more rapid evolution of the source population than the diffuse
component is found. No evidence for variability is found between the
Chandra and previous observations.

\end{abstract}


\section{Introduction}

Though galaxies were once thought of as `Island Universes', evolving
slowly in complete isolation, this is now known not to be the
case. Galaxies interact in a wide variety of ways with their
environment, and collisions and mergers of galaxies are now believed
to be one of the dominant mechanisms in galactic evolution (Schweizer
1989). There are probably very few galaxies today that were not shaped
by interactions or even outright mergers. Indeed, Toomre's (1977)
hypothesis, whereby elliptical galaxies might be formed from the
merger of two disc galaxies, is now generally accepted, such behaviour
having been modelled in many N-body simulations of mergers
(e.g. Toomre \& Toomre 1972; Barnes 1988). During such an encounter,
the conversion of orbital to internal energy causes the two progenitor
disks to sink together and coalesce violently into a centrally
condensed system. The `Toomre sequence' (Toomre 1977) represents
probably the best examples of nearby ongoing mergers, from disk-disk
systems to near-elliptical remnants.

One of these is the famous binary interacting system NGC4676A/B (also
Arp242). First described by Vorontsov-Vel'yaminov (1957) as a pair of
``playing mice'', ever since, the name `the Mice' has become
universally accepted. It consists of two distinct spiral galaxies (the
edge-on NGC4676A to the north, and the more face-on NGC4676B to the
south) with long tidal tails, and the presence of these, along with an
obvious bridge between the two galaxies and an increase in IR activity
compared to typical field galaxies, indicates that the galaxies have
begun to interact, and have passed one another. It is one of the
original systems presented by Toomre \& Toomre (1972) as a classic
example of a pair of galaxies undergoing tidal interaction. It lies
second in the proposed evolutionary sequences of both Toomre (1977)
$-$ the {\em Toomre sequence}, and of Hibbard \& van Gorkom (1996). It
is also placed second in the first study of the X-ray properties of an
evolutionary sample of merging galaxies; the ROSAT work of Read \&
Ponman (1998) (hereafter RP98).

The systematic velocity of the system appears well-defined in several
different regions of the electromagnetic spectrum (e.g. Stockton 1974;
Smith \& Higdon 1994; Hibbard \& van Gorkom 1996; Sotnikova \&
Reshetnikov 1998), and this converts, assuming \ho\ = 75\,km s$^{-1}$
Mpc$^{-1}$, to the distance to the Mice used in this paper;
88\,Mpc. It is generally agreed, using both kinematical work
(e.g. Toomre \& Toomre 1972; Mihos, Bothun \& Richstone 1993) and
multiwavelength observations (e.g. Hibbard \& van Gorkom 1996), that
both galaxies are in the throes of a prograde encounter, and have
their northern edges moving away from us. Hence, NGC4676A's tail is on
the very furthest side, swinging away from us (the tail has a
systematic velocity almost 300\,km s$^{-1}$ faster than the NGC4676A
nucleus; Sotnikova \& Reshetnikov 1998), and NGC4676B is rotating
clockwise, with its northeastern edge lying closest to us.

Both galaxies appear (e.g. from the optical data of Schombert, Wallin
\& Struck-Marcell 1990) to have shapes and colours consistent with
those of early-type spirals, though the disc regions are strongly
distorted or absent. Both tails, although bluer than the galaxies'
central colours, are in agreement with the colours of outer-disc
regions. The northern tail has a very high surface brightness
($\mu_{B} = 23.0-23.5$\,per square arcsecond) when compared with
similar features in other galaxies, whereas the southern tail is
roughly 1 magnitude per square arcsecond fainter (Schombert \etal\
1990). The tails account for 16\% of the total $H{\alpha}$ emission,
are quite luminous, containing one-third of the total R-band
luminosity of the system, and have a high atomic gas content (Hibbard
\& van Gorkom 1996). Local $H{\alpha}$ maxima are also observed in the
northern tail (Sotnikova \& Reshetnikov 1998).  The northern galaxy
appears to exhibit a 6.6$h^{-1}$\,kpc plume of $H{\alpha}$ along its
minor axis, and the southern galaxy possesses an ionized gas bar, as
produced in Barnes \& Hernquist's (1991, 1996) merger simulations,
offset with respect to the stellar bar (Hibbard \& van Gorkom
1996). Angular momentum transfer between the two bars is able to force
large amounts of gas towards the galactic centre.  Recent mapping of
the CO emission in the Mice (Yun \& Hibbard 2001) detect a compact
(\ltsim 2\,kpc), disklike or ringlike complex, perhaps accounting for
20\% of the total nuclear mass, centered on the inner stellar disk of
NGC4676A, with kinematics consistent with simple rotation. Within
NGC4676B, the CO emission is far less bright and occurs along the 7
kpc stellar bar.

A very relevant work is the set of papers dealing with the Chandra
X-ray emission from the Antennae (Fabbiano, Zezas \& Murray 2001,
Zezas \etal\ 2002, Zezas \& Fabbiano 2002), a very similar system to
the Mice, in that it involves the merger of two equal-sized gas rich
spiral galaxies. The Antennae however lie at a later evolutionary
epoch, in that the two galactic disks have begun to interact violently
with each other. Also, the Antennae lie at a much closer distance than
the Mice, aiding the detection of low luminosity sources, and
decreasing the effects of source confusion. Various aspects of the two
systems' X-ray emission are compared throughout this paper.

In this paper results of a $\sim$30.5\,ks Chandra ACIS-S observation
of the Mice are presented. The fundamental advantage of Chandra over
any other previous or current X-ray mission, is its excellent spatial
resolution. With a spatial resolution some 10 times better than the
ROSAT HRI and XMM-Newton, it is possible to resolve emission on scales
down to $\sim$210\,pc ($\approx 70\%$ encircled energy PSF [on-axis at
$<$6\,keV]) at the distance of the Mice. This allows the removal of
point source emission, and the analysis of any unresolved, perhaps
diffuse emission. In addition, the spectral resolution of Chandra is
comparable to that of ASCA, and far better than that of the ROSAT PSPC
(the ROSAT HRI note, having essentially no spectral resolution
whatsoever), allowing us to study the X-ray spectral properties of the
sources and emission regions seen, providing additional information as
to their nature.

The X-ray observations of the Mice prior to Chandra are described in
the following subsection. Section~2 describes the Chandra observations
and the data reduction techniques used. Discussion of the spatial,
spectral and temporal properties of the source and diffuse emission
components follow in Section~3, and in Section~4, the conclusions are
presented.

\subsection{Previous X-ray Observations}

The Mice have only previously been observed in X-rays with ROSAT. The
PSPC observations were described in RP98, and show rather an amorphous
X-ray structure, with only one source detected, lying between the two
galactic nuclei. Little could be said as regards the spectral
properties of the source. The general form of the PSPC emission
however, seemed to follow the optical `heads' of the two galaxies,
running essentially north-south, with some tentative evidence for
extension in the east-west direction, especially around
NGC4676A. Though only a small number of counts were obtained, the
fitted temperature to the diffuse emission spectrum was seen to be in
good agreement with fitted temperatures of nearby, known, starburst
winds (Heckman 1993; Read, Ponman \& Strickland 1997), suggesting a
starburst origin for the diffuse emission.

\begin{figure}
\vspace{8cm} 
\includegraphics{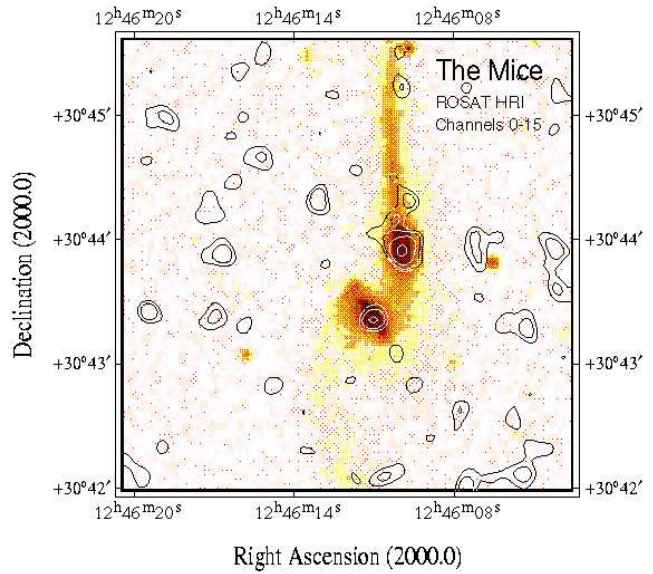} 
\caption{Contours of ROSAT HRI emission over the full channel range
superimposed on an optical Digital Sky Survey (DSS) image of the
Mice. The 2\arcs\ resolution X-ray image has been smoothed with a
Gaussian of FWHM 9.4\arcs. Contours are at values of 2, 3, 5 and
9$\sigma$ ($\sigma$ being 52.4\,counts/arcmin$^{2}$) above the
background (157.3\,counts/arcmin$^{2}$).  }
\label{fig1}
\end{figure}

The Mice were also observed with the ROSAT HRI. The previously
unpublished results of these observations are presented here, as it
serves both as a useful introduction to the system and its X-ray
structure, and as a stepping-stone from the (relatively)
low-resolution of the ROSAT PSPC to the high resolution of Chandra.

The 33.4\,ks of ROSAT HRI data (ID 601091), taken at the end
of 1997, were initially screened for good time intervals longer than
10\,s. No further selection on low background times was made, as we
were mainly interested in point-like sources (which the HRI is far
better suited to), and for this purpose, the data were still
photon-limited. Source detection and position determination was then
performed over the full field of view with the EXSAS local detect, map
detect and maximum likelihood algorithms (Zimmermann \etal\ 1994),
using images of pixel size 5\arcs.

Fig.~1 shows smoothed contours of ROSAT HRI emission (over the full
channel range) from the Mice superimposed on an optical Digital Sky
Survey (DSS) image. Though there are very few counts, it is clearly
seen that the HRI is able to resolve the PSPC emission into two
distinct sources at the positions of the two galaxies. These were the
only two sources detected in the vicinity of the optical galaxies. The
northern source is detected (with 22 source counts at
$\alpha[2000.0]=12^{\rm h}46^{\rm m}9.95^{\rm s}$
$\delta[2000.0]=+30^{\circ}43\arcm 54.3\arcs $) at a significance of
5.9$\sigma$ with a (0.1$-$2.4\,keV) count rate of
(6.8$\pm$1.7)$\times10^{-4}$\,counts s$^{-1}$, the southern source
(with 10 source counts at $\alpha[2000.0]=12^{\rm h}46^{\rm
m}11.04^{\rm s}$ $\delta[2000.0]=+30^{\circ}43\arcm 21.4\arcs $) at
3.5$\sigma$ with a count rate of (3.1$\pm$1.2)$\times10^{-4}$\,counts
s$^{-1}$. Errors on the HRI positions are 1.4$-$1.6\arcs. Further,
perhaps extended emission is suggested to the north and east of
NGC4676A, and this is borne out by a low-significance extent within
the source-searching results to the northern source's emission.

\section{Chandra observations, data reduction and results}

The Mice were observed with Chandra on May 29th, 2001 for a total of
just over 30\,ks, with the back-illuminated ACIS-S3 CCD chip at the
focus (Observation ID: 2043). Data products, correcting for the motion
of the spacecraft and applying instrument calibrations, were produced
using the Standard Data Processing (SDP) system at the Chandra X-ray
Center (CXC). These products were then analysed using the CXC CIAO
software suite (version 2.2.1). A lightcurve extracted from a large
area over the entire observation was seen to be essentially constant
and consistent with a low-level rate. Consequently, no screening to
remove periods of high background flaring was performed.
 
\subsection{Overall X-ray structure}

Fig.\,2 (left) shows contours of adaptively smoothed (0.2$-$10\,keV)
Chandra ACIS-S X-ray emission from the field surrounding the Mice
system, superimposed on an image from the Hubble Space Telescope's
Advanced Camera for Surveys (ACS) instrument. The adaptive smoothing
of the X-ray emission attempts to adjust the smoothing kernel to
obtain a constant signal-to-noise ratio across the image.

Several things are immediately evident from the image. As suggested by
the HRI data, the emission is resolved into two main components
associated with the two galaxy `heads'. The factor 10 higher
resolution over the HRI however, is able to shed far more light on the
situation. The emission associated with NGC4676A is centered close to
the galaxy nucleus, and appears predominately diffuse and extended,
especially in the east-west direction (\ie\ along the minor axis of
the galaxy, NGC4676A being seen almost exactly edge on). The X-rays
within NGC4676B appear far more centrally concentrated at the nucleus,
though some evidence for extension is also seen again along NGC4676B's
minor axis (\ie\ from south-east to north-west). Orthogonal to this,
hotspots of X-ray emission are seen along the galaxy disk, equidistant
from the nucleus, especially to the south-west. Other interesting
features are seen associated with the tails, one source lying directly
in the northern tail, and a further source lying to the east of the
southern tail. No significant features are seen associated with the
more distant, off-image tail regions. One last interesting feature is
the source seen $\sim$40\arcs\ west of NGC4676A.

\begin{figure*}
\vspace{12cm} 
\includegraphics{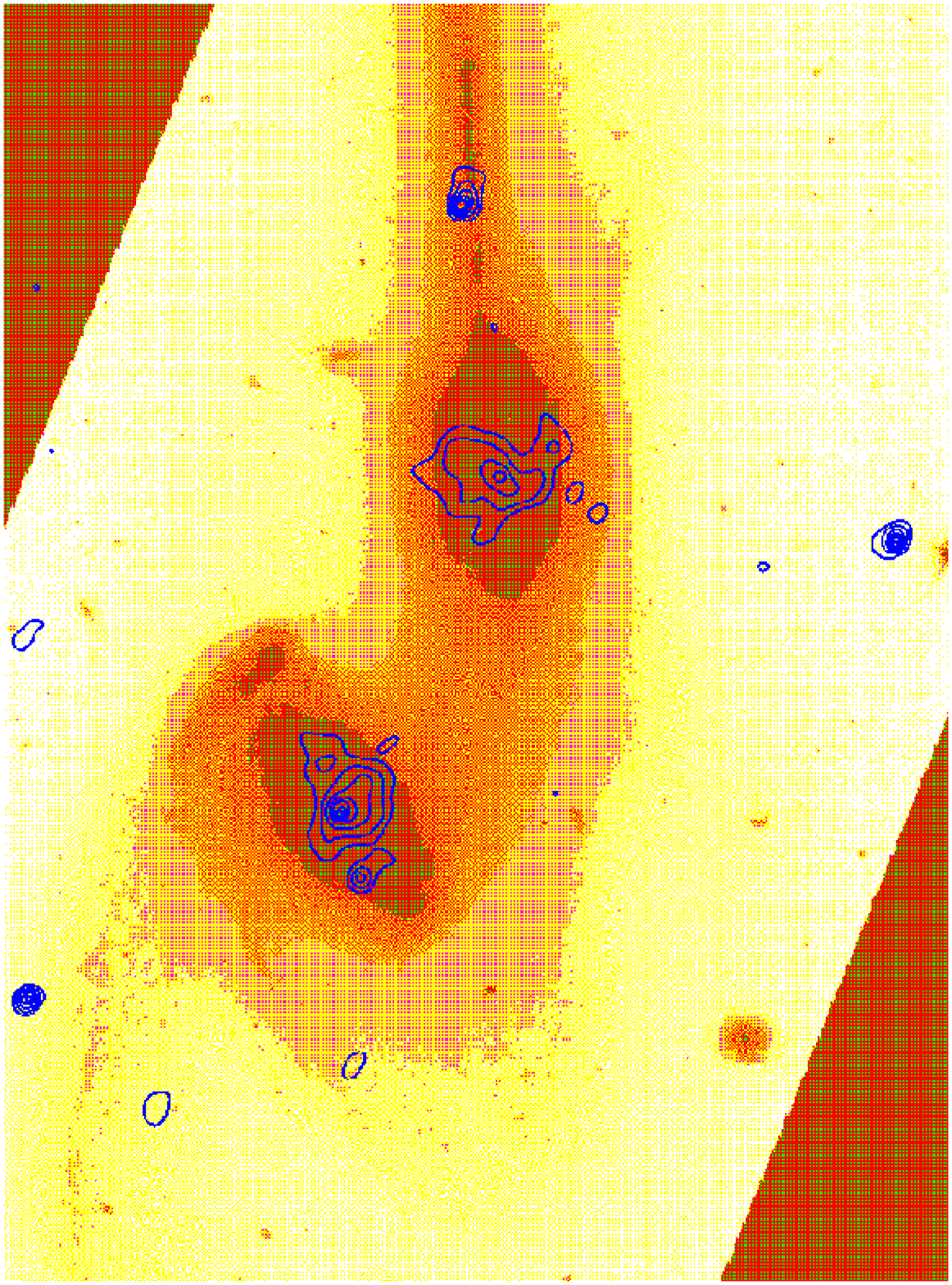} 
\includegraphics{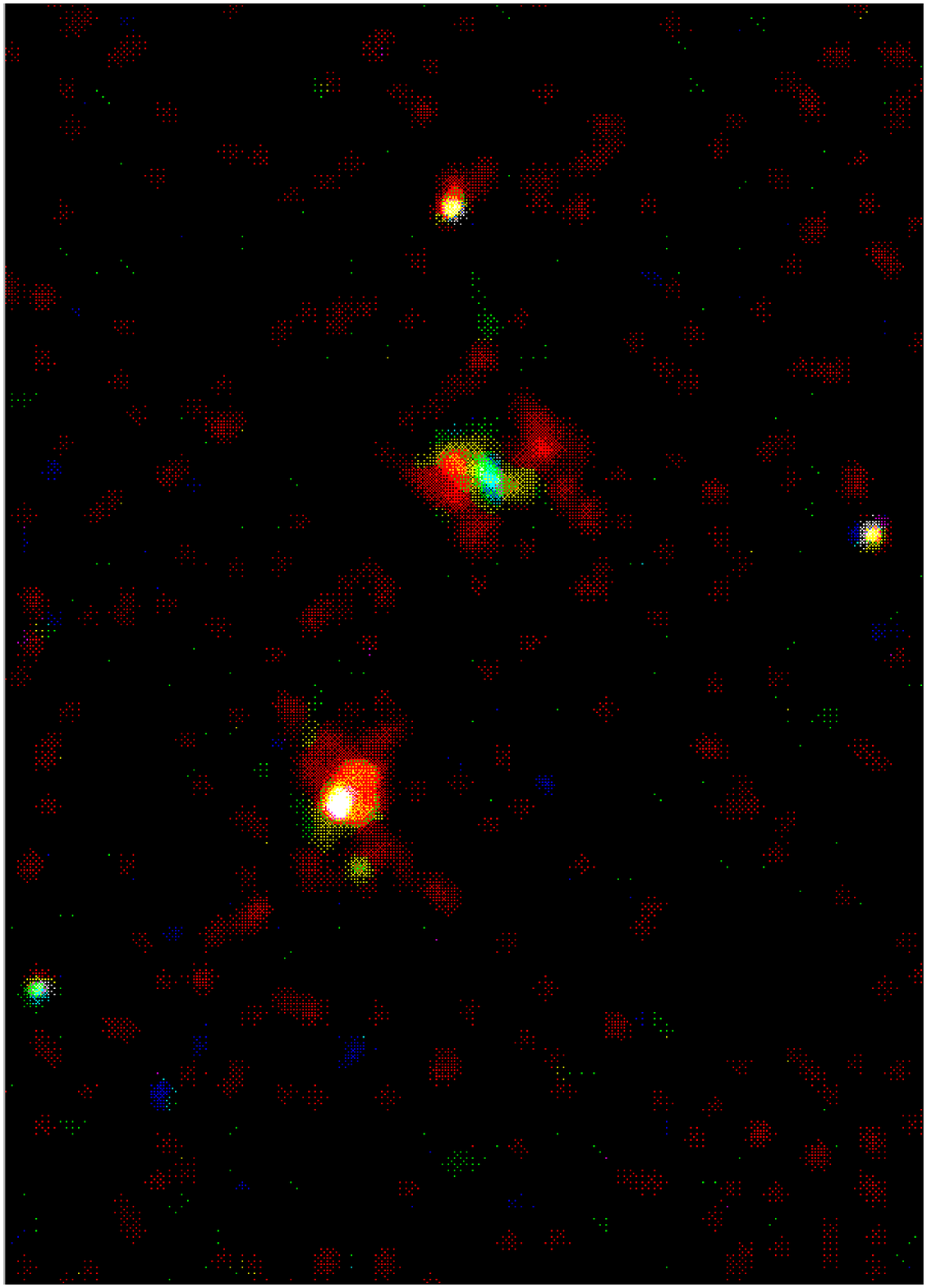} 
\caption{(Left) Contours of adaptively-smoothed (0.2$-$10\,keV)
Chandra ACIS-S X-ray emission from the field surrounding the Mice
system, superimposed on a Hubble Space Telescope ACS image. The X-ray
contours increase by factors of two. For scale, the X-ray sources at
the two galaxy `heads' lie $\approx$36\arcs\ apart. (Right) `True
colour' X-ray image of the Mice to the same scale. Red corresponds to
0.2$-$0.9\,keV, green to 0.9$-$2.5\,keV and blue to 2.5$-$10\,keV.}
\end{figure*}

Fig.\,2 (right) shows a true colour X-ray image, with red
corresponding to 0.2$-$0.9\,keV, green to 0.9$-$2.5\,keV and blue to
2.5$-$10\,keV. The individual adaptively-smoothed images were obtained
as described above. The image is to the same scale as Fig.\,2 (left),
and all the X-ray features are visible. The northern head shows a
medium energy nucleus, with cooler, complex, structured emission
surrounding. The southern head is brighter and hotter than the
surrounding emission which again is cool (white indicates strong
emission in all three bands). The nearby feature SW of the southern
head appears soft, as does the feature in the northern tail. The
feature just east of the southern tail and the source west of the
system appear somewhat harder.

\subsection{Point sources: spatial and spectral properties}

The CIAO tool {\em wavdetect} was used to search for point-like
sources, on scales from 1$-$16 pixels (0.5$-$8\arcs). A total of 6
sources were detected in the 0.3$-$10\,keV band within or close to the
optical confines of the galaxies (whether the galaxy `heads' or
tails), and their X-ray properties are summarized in Table~1. Sources
are detected (see Fig.\,3) at the positions of the two galaxies (A and
B), and a further source (B2) is detected just south-west of
NGC4676B. A source (N) is detected within the northern tail, and a
source (S) is detected close to the eastern edge of the southern
tail. Finally, a source (W) is detected west of the the main body of
the system. The X-ray properties given in Table~1 are as follows;
Right Ascension and Declination (2000.0) are given in cols.\,2 and 3,
together with the positional error (in arcseconds) in col.\,4. Net
source counts (plus errors) are given in col.\,5, and the source
significance is given in col.\,6. Hardness ratios (plus errors) are
given in col.\,7. These were calculated by performing additional
detection runs (using again scales of 1$-$16 pixels) in a soft
(0.3-2\,keV) and a hard (2-10\,keV) band. The tabulated values are
$(H-S)/(H+S)$, $H$ being the net source counts in the hard band, $S$
being the net source counts in the soft band. Finally, (0.3$-$10\,keV)
X-ray emitted and intrinsic (\ie\ corrected for absorption)
luminosities (obtained as described in the spectral fitting section)
are given in cols.\,8 and 9.

\begin{table*}
\caption[]{Sources detected by {\em wavdetect} in the 0.3$-$10\,keV
band within or close to the optical confines of the Mice. Columns are
described in the text. Luminosities assume (except for source~B; see
text) a photon index 1.5 power law plus Galactic absorption, and a
distance of 88\,Mpc.}
\begin{tabular}{lcccrrccc}
\noalign{\smallskip}
\hline
Src. & RA          &Dec.          &Pos.err.& Counts(err)     & Sig. & HR & \multicolumn{2}{c}{$L_{X}$ (0.3$-$10\,keV)} \\ 
     &\multicolumn{2}{c}{(2000.0)}&(arcsec)&                 &      &    & \multicolumn{2}{c}{($10^{39}$\,erg s$^{-1}$)} \\ 
     &             &              &        &                 &      &    & (emitted) & (intrinsic) \\ \hline
A    & 12 46 10.09 & +30 43 55.8  & 0.22   & 26.52$\pm$ 5.29 &10.64 & -0.45$\pm$0.27 & 11.92 & 12.17 \\
B    & 12 46 11.23 & +30 43 22.0  & 0.12   & 92.98$\pm$ 9.80 &31.59 & -0.48$\pm$0.12 & 23.99 & 26.14 \\
B2   & 12 46 11.08 & +30 43 15.8  & 0.23   & 10.29$\pm$ 3.31 & 4.66 & -0.66$\pm$0.36 & 3.22 & 3.30 \\
N    & 12 46 10.36 & +30 44 22.2  & 0.14   & 27.25$\pm$ 5.29 &12.25 & -0.59$\pm$0.21 & 8.72 & 8.91 \\
S    & 12 46 13.62 & +30 43 03.3  & 0.13   & 17.36$\pm$ 4.24 & 7.96 & -0.48$\pm$0.36 & 4.98 & 5.04 \\
W    & 12 46 07.05 & +30 43 49.4  & 0.14   & 23.26$\pm$ 4.90 &10.47 & -0.44$\pm$0.23 & 7.82 & 7.98 \\
\noalign{\smallskip}
\hline
\end{tabular}
\end{table*}

Source spectra were extracted in the range 0.3$-$10\,keV at the exact
positions given by the 0.3$-$10\,keV detection analysis. The regions
output by the detection routines, defined to include as many of the
source photons as possible, but minimizing the background
contamination, were invariably near-circles of radius $\sim5$ pixels
(partly due to the Mice only occupying the very centre of the ACIS-S3
chip), and consequently, extraction circles of radius 5 pixels
(2.5\arcs) were used for all the sources in Table~1. Background
regions were source-free circular annuli surrounding each source (in
order to minimize effects related to the spatial variations of the CCD
response), though regions of apparent diffuse emission were also
avoided in the extraction of background regions; in the cases of A, B
and B2, background extraction annuli of inner-to-outer radii
15$-$22.5\arcs\ were used. In the cases of N, S and W, background
extraction annuli of 2.5$-$12.5\arcs\ were used. All source and
background extraction regions are shown in Fig.\,3.

\begin{figure}
\vspace{12cm} 
\includegraphics{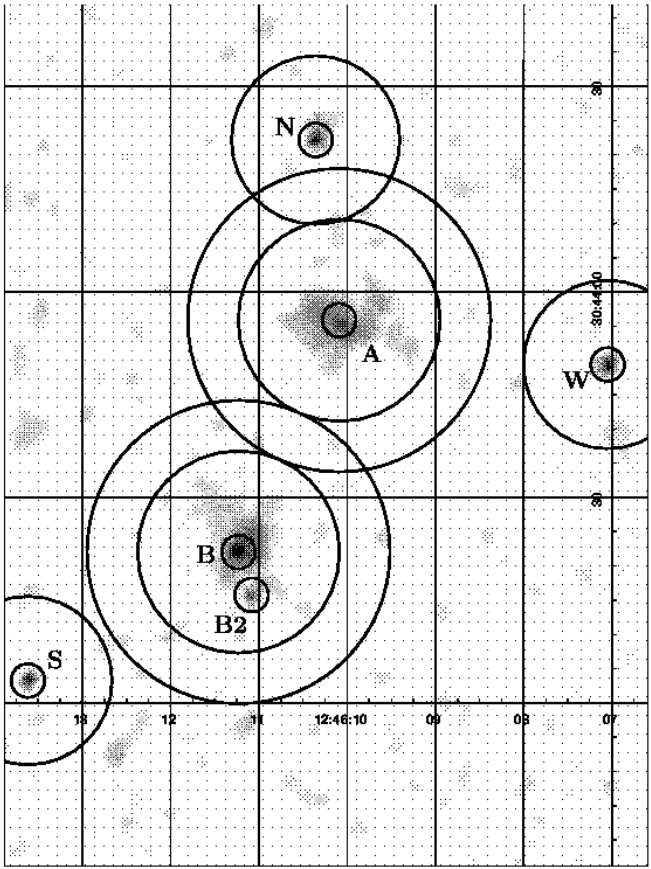} 
\caption{Adaptively-smoothed (0.2$-$10\,keV) ACIS-S image (scale as
for Fig.\,2) labelled with sources detected in the 0.3$-$10\,keV
band. Circles show the source, diffuse and background regions used in
the extraction of the spectra (see text).}
\end{figure}

ACIS spectra were extracted using Pulse Invariant (PI) data values,
and were binned together to give a minimum of 10 counts per bin after
background subtraction. Hence $\chi^{2}$ statistics could be
used. Response matrices and ancillary response matrices were created
for each spectrum, using the latest calibration files available at the
time of writing.

Standard spectral models were fit to the spectral data using the CIAO
spectral fitting software. Events above 7\,keV (of which there were
very few) and below 0.3\,keV were excluded from the fitting on the
grounds of uncertainties in the energy calibration. It is now known
that there has been a continuous degradation in the ACIS QE since
launch. A number of methods now exist within the community to correct
for this. These include the release of an XSPEC model (ACISABS) to
account for this degradation, and the existence of software (corrarf)
to correct the ancillary response files. Both methods have been used
here in the spectral fitting, and very similar results were
obtained. In both cases, the time since launch of the observations
(here, 678 days) is used in the correction. Although the calibration
at energies below 1.0\,keV is believed to be uncertain, data in this
range were kept, as the statistical error on these data points is
still greater than the errors due to the uncertainties in the
calibration.

Two models, one incorporating absorption fixed at the value out of our
Galaxy (1.28$\times10^{20}$\,cm$^{-2}$) and a 5\,keV {\em mekal}
thermal plasma, the other incorporating absorption (again, fixed) and
a power-law of photon index 1.5 were fit to the data. Only for
source~B were there sufficient counts to let the model parameters fit
freely. A good fit (reduced $\chi^{2}$\ltsim1) was obtained using a
power-law model of photon index 1.52 and an absorbing column of
4.19$\times10^{20}$\,cm$^{-2}$. Fig.\,4 shows the 99\%, 90\% and 68\%
confidence contours in the photon index-absorption column plane for
the power-law fit to the spectrum of source~B. The luminosities quoted
in Table~1 for source~B assume this best-fitted model, while for the
other sources, the model assumed is of fixed (Galactic) absorption
plus a power-law of photon index 1.5.

\begin{figure}
\vspace{7cm} 
\includegraphics{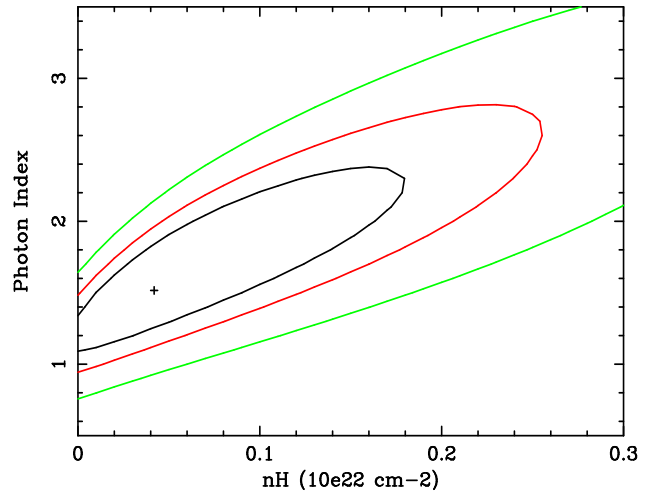} 
\caption{99\%, 90\% and 68\% confidence contours in the photon
index-absorption column plane for the power-law fit to the spectrum of
source~B.}
\end{figure}

\subsection{Residual emission: spatial and spectral properties}

The existence of residual, likely diffuse emission, surrounding the
two head sources A and B is very evident in the figures. Spectra from
these regions of apparent diffuse emission were extracted in the range
0.3$-$10\,keV. An annulus of 2.5$-$15\arcs\ centred on source~A was
used to extract a spectrum of the `diffuse' emission associated with
NGC4676A (hereafter `diffA'). An identical annulus around source~B
(excepting a 2.5\arcs\ circular region surrounding source~B2) was used
to extract a spectrum of the emission around NGC4676B
(`diffB'). Background regions were as for sources~A and B. These
diffuse and background extraction regions are shown in Fig.\,3. 

The spectral fitting was performed as for the point sources, using the
same models, and using both methods to correct for the degradation in
the ACIS QE. With $\approx$80$-$100 non-background counts in each
spectrum (see Table\,2), it was possible to place some constraints on
the spectral properties of the residual emission. In both cases, diffA
and diffB, an absorption plus power-law model was unable to fit the
data satisfactorily. Only reduced $\chi^{2}$'s of 1.5 (diffA) and 1.4
(diffB) were attainable using power-law models. Thermal models proved
much better, and the best thermal fits (again using an absorption plus
{\em mekal} model) are summarized in Table~2; the absorbing column,
the fitted temperature and metallicity (an `F' indicating a fixed
value), the reduced $\chi^{2}$, and the emitted and intrinsic (\ie\
absorption-corrected) X-ray luminosity is given. Essentially identical
temperatures ($\approx0.5\pm0.2$\,keV) are obtained for both diffuse
emission features, and the data plus best fit models are shown in
Fig.\,5.

\begin{table*}
\caption[]{Best results of fitting thermal models to the spectra of
residual emission around sources~A and B. Luminosities assume a
distance of 88\,Mpc (see text).}
\begin{tabular}{lccccccc}
\noalign{\smallskip}
\hline

Diff. & Counts(err)    & \nh\                   & $kT$  & $Z$     & $\chi^{2}$ & \multicolumn{2}{c}{$L_{X}$ (0.3$-$10\,keV)} \\
Src.  &                & ($10^{20}$\,cm$^{-2}$) & (keV) & (solar) & (red.)     & \multicolumn{2}{c}{($10^{39}$\,erg s$^{-1}$)} \\ 
      &                &                        &       &         &            & (emitted) & (intrinsic) \\ \hline 
diffA & 103.4$\pm$12.7 & 1.28(F) & 0.50$^{+0.18}_{-0.13}$ & 0.3(F) & 1.20 & 10.13 & 10.86 \\
diffB & 80.55$\pm$11.9 & 1.28(F) & 0.46$^{+0.20}_{-0.12}$ & 0.3(F) & 0.45 & 7.34 & 7.89 \\

\noalign{\smallskip}
\hline
\end{tabular}
\end{table*}

\begin{figure}
\vspace{15cm} 
\includegraphics{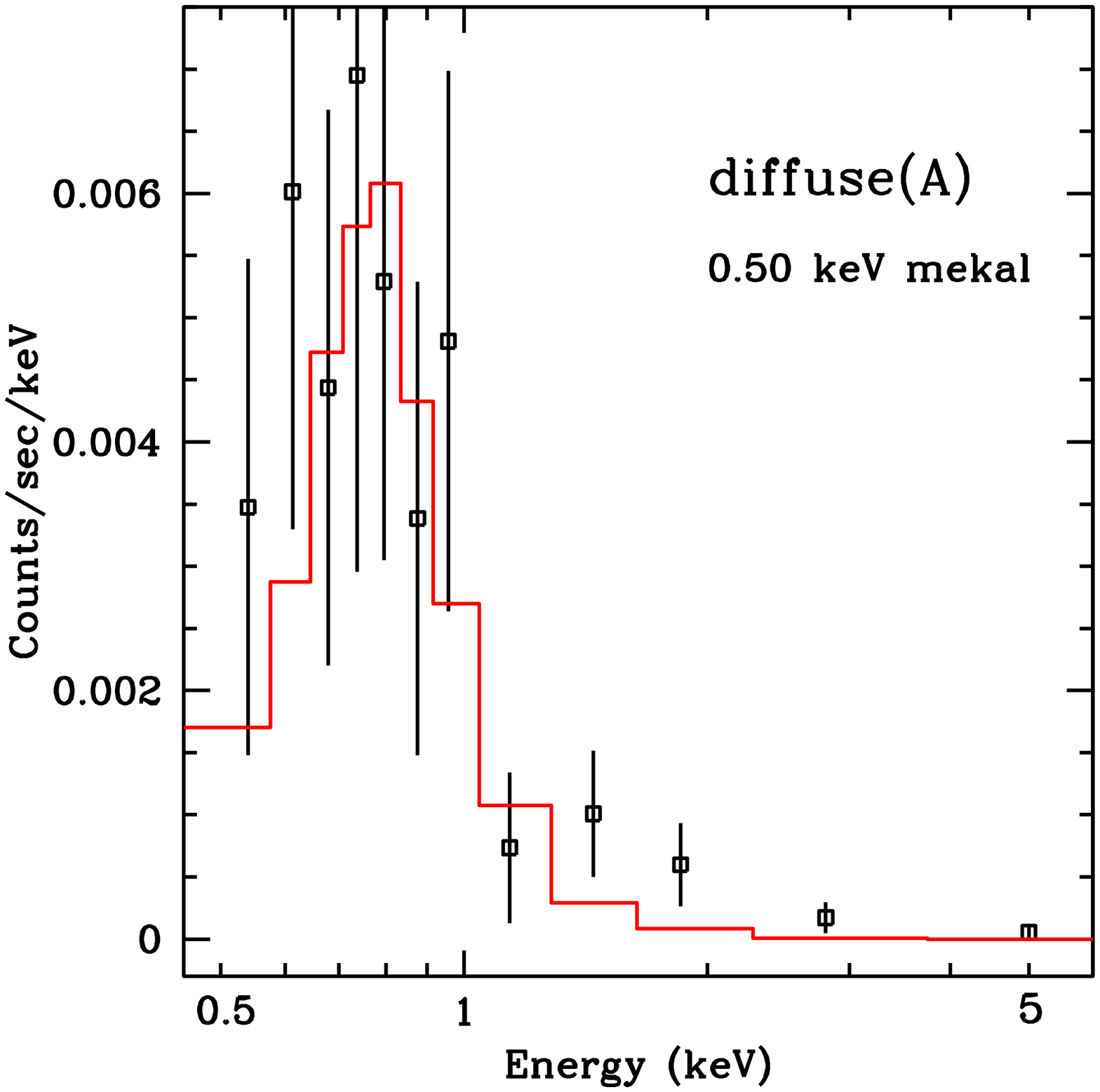} 
\includegraphics{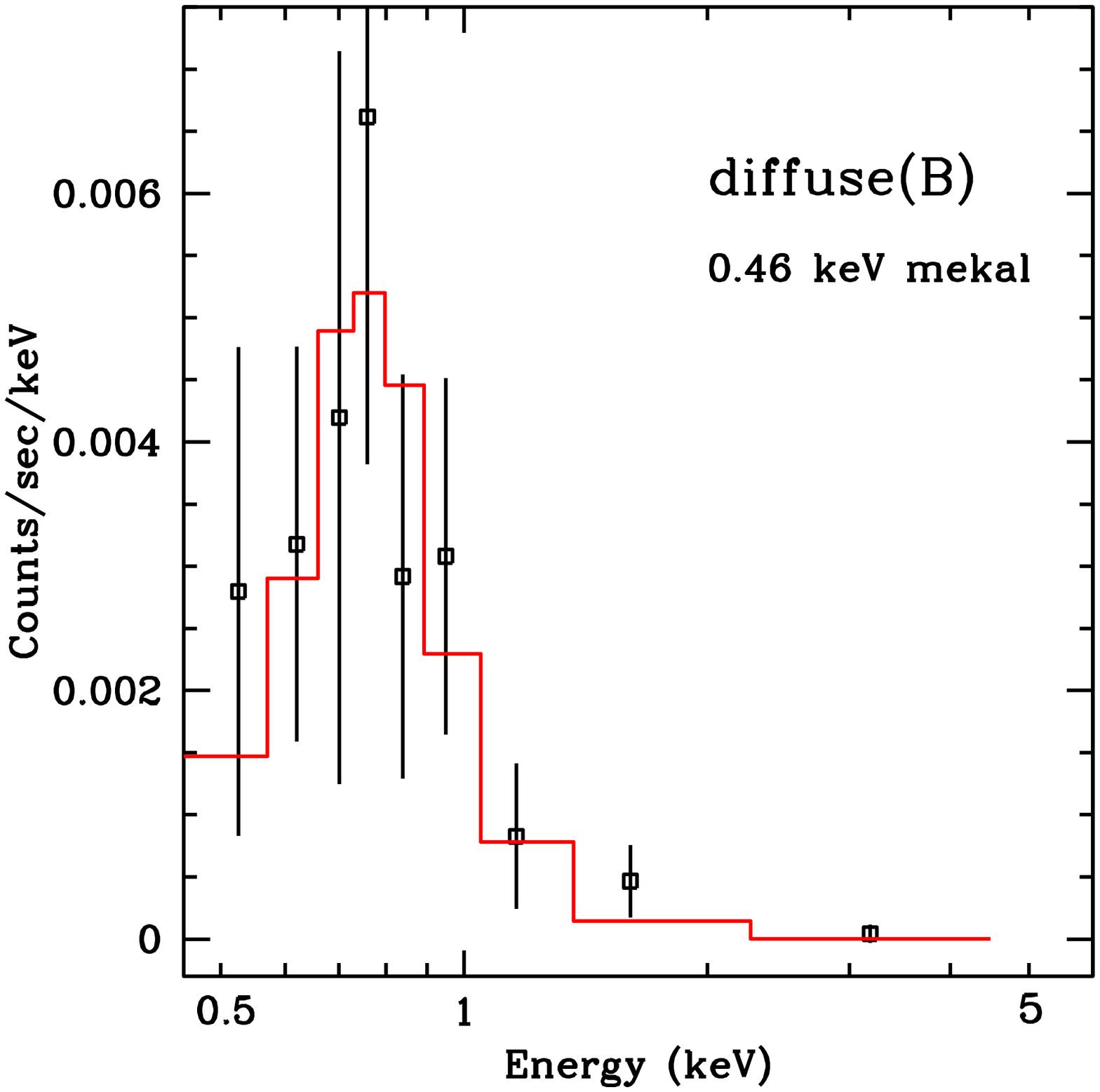} 
\caption{Data (points) plus best-fit thermal mekal models (lines) for the residual emission 
around source~A (top) and source~B (bottom). 
}
\end{figure}

\subsection{Temporal properties}

As we have two observations of the Mice where one can distinguish
discrete emission regions (the first being the ROSAT HRI observations,
described in Sect.\,1.1), some constraints can be placed on the
temporal properties of those regions. X-ray luminosities in the
0.3$-$10\,keV band have been calculated for the ROSAT HRI count rates,
using the ACIS-S best-fit models to the source~A and B
spectra. Comparison of the HRI and ACIS-S PSFs indicates that the HRI
would essentially `see' source~A and most of the residual emission
around source~A (diffA) as one point source. Similarly, the HRI would
detect the emission from sources~B, B2 and from diffB as one
source. One can therefore only study the temporal variation of the
emission from the `heads' of the two galaxies; NGC4676A (where the
total ACIS-S (intrinsic) $L_{X}$ is the sum of the X-ray luminosities
of source~A and diffA) and NGC4676B (the total $L_{X}$ being the sum
for sources~B, B2 and diffB). In Fig.\,6, these two crude X-ray
lightcurves are shown. Errors on the luminosities are taken as the
statistical errors on the count rates (the errors on the HRI
luminosities are very large, due to there being very few counts, and a
higher background level). For the southern galaxy NGC4676B, the level
of the emission appears not to have varied at all between the two
observations. If one uses a power-law model to calculate the HRI
luminosity of NGC4676A (the upper 1997 point in the left-hand panel of
Fig.\,6), then it appears that some variation may have occurred (at
least at the $\sim2\sigma$ level). However, re-calculating an
intrinsic (0.3$-$10\,keV) X-ray luminosity for the NGC4676A ROSAT HRI
count rate, but using instead the best-fit model to the diffA
spectrum, results in a value (the lower 1997 point in the left-hand
panel of Fig.\,6) very in accordance with the ACIS-S value. The true
HRI luminosity lies somewhere between these two points, and one can
conclude therefore that there are no significant signs of any X-ray
variations between the two observations.

\begin{figure}
\vspace{7.5cm} 
\includegraphics{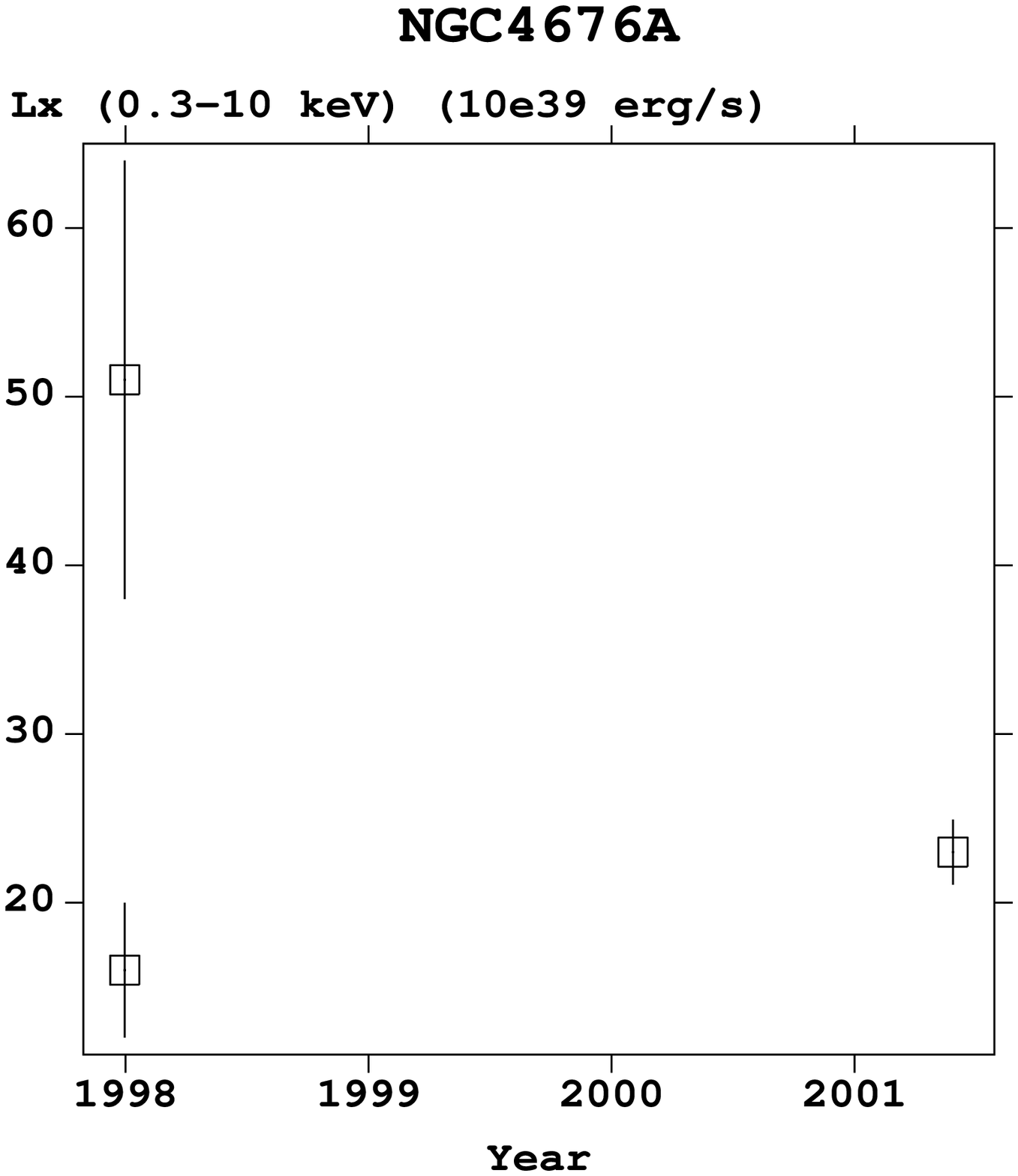}
\includegraphics{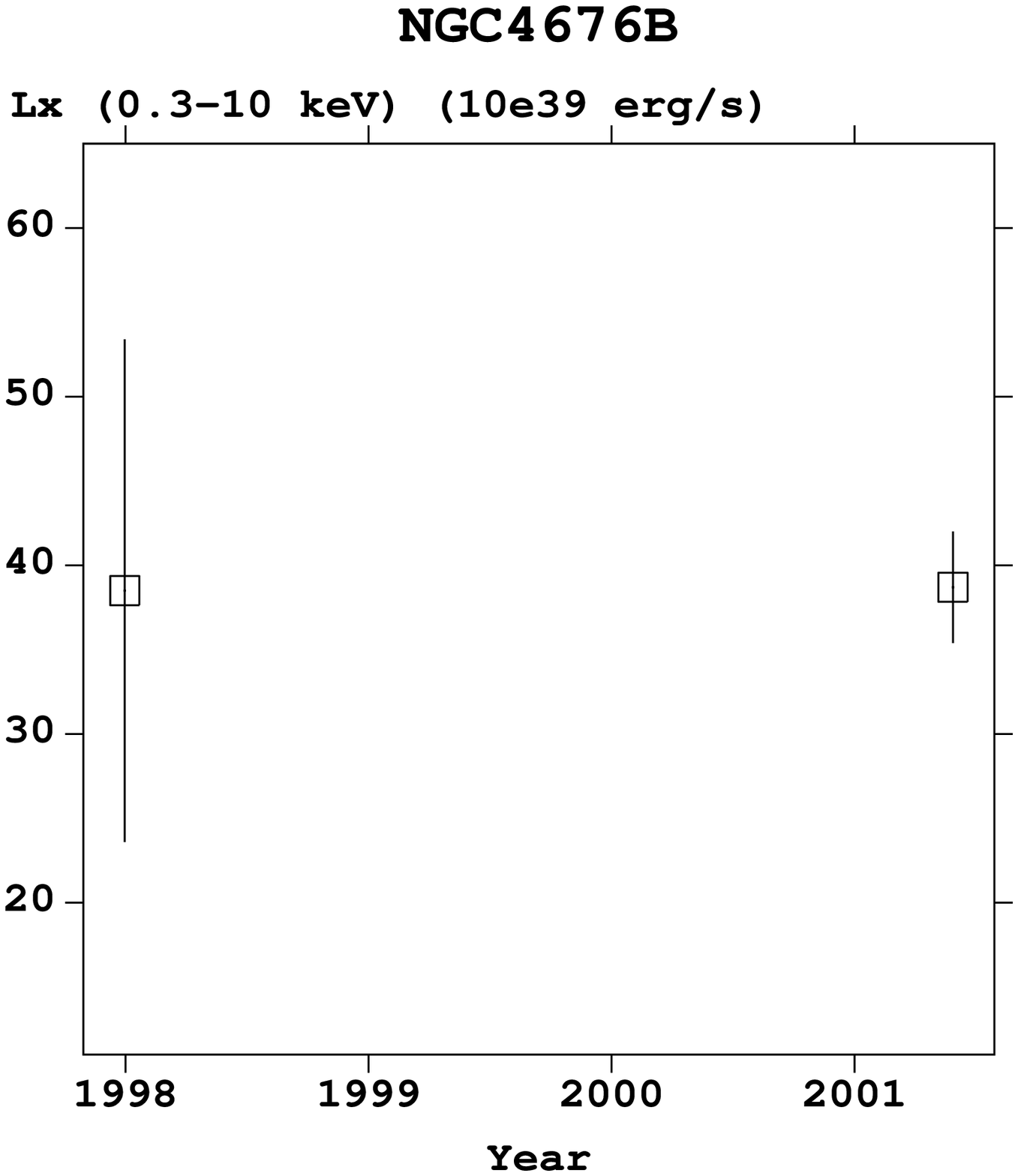}
\caption{Crude X-ray (intrinsic) luminosity lightcurves of the
emission regions associated with the `heads' of the two galaxies;
NGC4676A (left) and NGC4676B (right). The two 1997 NGC4676A points
refer to the usage of two different models (see text). }
\end{figure}

\section{Discussion}

\subsection{Point sources}

With the Chandra ACIS-S instrument, it is possible to resolve more
clearly the source populations within the Mice, that were suggested by
the ROSAT HRI observations.

Complex emission regions are seen at the centres of the two galaxies,
surrounded by what appears to be residual, diffuse X-ray emission, and
further high-luminosity sources are detected close to the southern
nucleus and within the tidal tails.

Recent mapping of the 2.6\,mm CO $J=1\rightarrow 0$ emission in the Mice
(Yun \& Hibbard 2001) has revealed large amounts of molecular gas. A
compact molecular complex is seen well centred on the inner stellar
disk of NGC4676A, forming a disk- or ring-like structure, seen within
5\dgr$-$10\dgr of being edge-on, with a deconvolved size of 1.8\,kpc
in radius and a thickness (FWHM) of 250\,pc. The gas kinematics
observed are consistent purely with rotation. A total molecular gas
mass of 5.5$\times 10^{9}$\,\mdot\ is detected (about twice as much as
in our Galaxy), making up a significant fraction (20\%) of the
total mass in the nuclear region. This is a very high fraction for
ordinary disk galaxies (Young \& Scoville 1991).

In NGC4676B, the situation is different in that only weak CO emission
is detected, and then only along the 7\,kpc stellar bar. Large peaks
in the CO emission are seen in NGC4676B at the ends of the bar, as is
often seen in other barred galaxies. The CO kinematics are consistent
with solid body rotation of the bar, and the molecular gas fraction is
seen to be about 7\%, which is more typical of undisturbed disk
galaxies.

The fact that several smaller CO clumps are seen in the bridging
region between the two Mice `heads' indicates that the disruption of
the inner disks has begun within the Mice. This, along with the
obvious morphological indicators of interaction, and the fact that an
increase in far-infrared luminosity is observed within the system, is
very suggestive of strong starburst activity having begun within the
one or both of the galactic nuclei. The Mice have $L_{FIR}/L_{B}$
values more like those of field starburst galaxies, than of normal
galaxies, and a large value of the far-infrared dust colour
temperature $S_{60}/S_{100}$, a diagnostic seen to increase with
interaction strength (Telesco \etal\ 1988), is also observed within
this system (RP98).

Significant X-ray emission is seen at the centres of the two
galaxies. Dealing first with source at centre of NGC4676B, source~B,
it is possible to place some constraints as to the spectral properties
of the source, in that an absorbed power-law is seen to fit the data
very well. This is not too constraining however, as the fit is not
significantly better than with an absorbed
(9.9$\times10^{20}$\,cm$^{-2}$) fixed-5\,keV mekal model. Though it is
difficult to say whether the emission at the centre of NGC4676B is
thermal or not, it appears to be rather absorbed, a fact not too
surprising, given the inclined orientation nature of the southern
galaxy.

Fig.\,7 shows white contours of 0.2$-$10\,keV ACIS-S X-ray emission
(as per Fig.\,2a) superimposed on Yun \& Hibbard's (2001) figure of
velocity-integrated CO (black contours) and $H{\alpha}$ (greyscale).
Fig.\,8 details the $H{\alpha}$ emission and shows contours of
$H{\alpha}$ (increasing by factors of 1.7783), from Hibbard \& van
Gorkom (1996) (and reproduced in Yun \& Hibbard (2001)), superimposed
on the 0.2$-$10\,keV ACIS-S X-ray emission (greyscale).

\begin{figure}
\vspace{10.5cm} 
\includegraphics{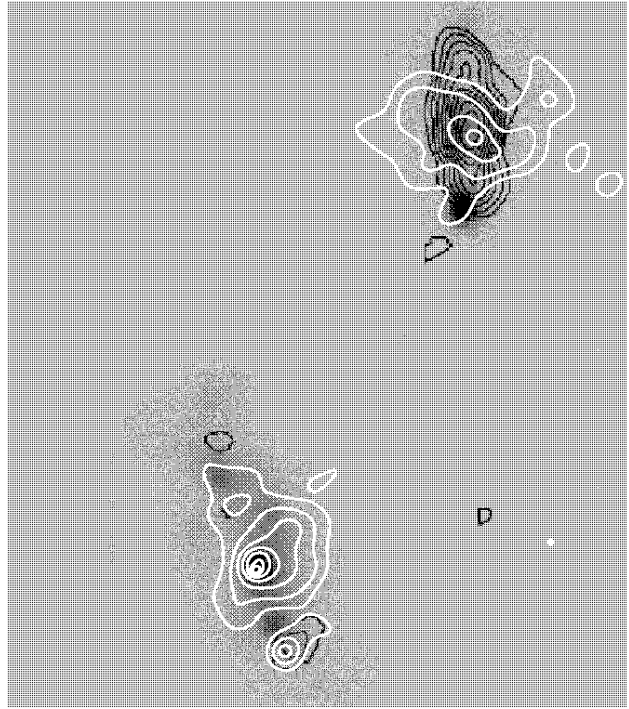} 
\caption{ 
0.2$-$10\,keV ACIS-S X-ray emission (white contours, as per Fig.\,2a)
superimposed on the Yun \& Hibbard (2001) figure of
velocity-integrated CO (1$-$0) maps (black contours) and $H{\alpha}$
(greyscale).
}
\end{figure}

\begin{figure}
\vspace{10.5cm} 
\includegraphics{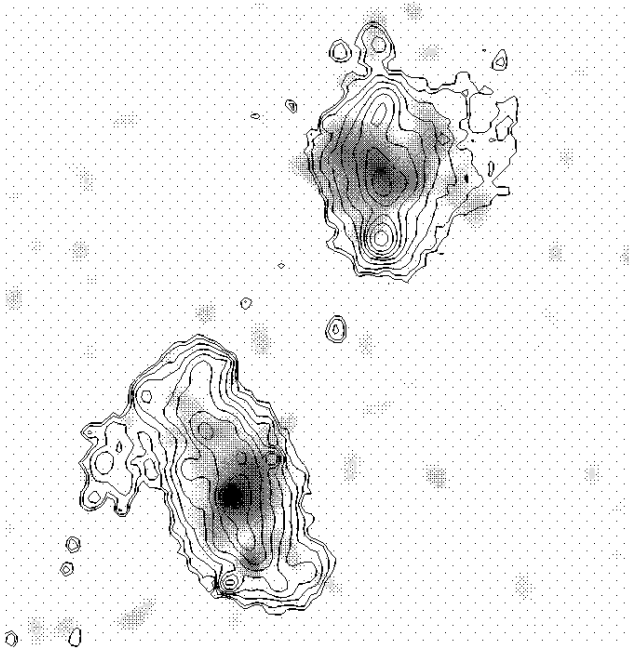} 
\caption{ 
Contours of $H{\alpha}$ (increasing by factors of 1.7783), from
Hibbard \& van Gorkom (1996), superimposed on 0.2$-$10\,keV ACIS-S
X-ray emission (greyscale).  }
\end{figure}

Source~B is very bright, appears very compact, and lies coincident
with the peak in the southern galaxy's $H{\alpha}$ emission, a site of
intense current star-formation. Little CO emission is seen here
however. In contrast, source~A, the X-ray source at the centre of
NGC4676A, is not as bright, nor as centrally compact, and appears
coincident with the peak in the CO emission disk, a site containing
large amounts of molecular gas. There is a peak in the $H{\alpha}$
emission at the position of source~A, indicating that a good deal of
star-formation is ongoing, though the main $H{\alpha}$ peak in the
northern galaxy lies some 13\arcs\ to the south, at the edge of the CO
disk. Yun \& Hibbard (2001) suggest that this near anticorrelation
between the CO and the $H{\alpha}$ emission in the northern galaxy is
due to extinction in the nearly edge-on disk. In fact the peak CO flux
corresponds to an \nh\ of $\approx6\times10^{22}$\,cm$^{-2}$, and this
may explain why source~A appears rather different to source~B in
X-rays, \ie\ we are not really seeing many of the X-rays (or indeed
the $H{\alpha}$ photons), because of the intervening absorbing
material in the disk. One cannot say anything using source A's
X-ray spectrum as regards absorption, but assuming again a 1.5
power-law model, but with an absorption suggested by the CO data, then
an intrinsic (0.3$-$10\,keV) X-ray luminosity of
2.1$\times10^{40}$\,erg s$^{-1}$ is obtained, more similar to that of
source~B. Sources A \& B may indeed be very similar. 

The fact that the X-ray emission at the centres of the two galaxies
appears spatially complex, with associated diffuse hot gas structures,
and significant amounts of molecular gas and $H{\alpha}$ emission,
suggests strongly (especially for source A) that the nuclear X-ray
sources are starburst-like in origin (as opposed to being
significantly dominated by AGN activity).

Such X-ray luminosities for the nuclear sources in the Mice are
certainly `Super-Eddington' (the Eddington limit for a 1\mdot\
accreting object is $\sim1.3\times10^{38}$\,erg s$^{-1}$), but these
nuclear sources may not be single sources, and we would need even
higher spatial resolution than that offered by Chandra to untangle any
complex source confusion.  It is interesting to note that the
luminosities of sources~A \& B are somewhat greater than those of the
two nuclear sources in the Antennae (Zezas \etal\ 2002). Though there
may be single massive accreting `active' sources at the centres of
these two galaxies, more likely is that these sources are collections
of supernovae and stellar winds plus some smaller accreting sources,
all embedded in diffuse hot gas $-$ a starburst. The fact that much
apparently diffuse X-ray emission is seen surrounding the nuclei
certainly supports this. This diffuse emission is discussed in the
next section.

Close by to the southern nucleus is source~B2, a rather soft source
according to its hardness ratio (Table~1) and its appearance in the`
true colour' image (Fig.~2b). It was not detected by ROSAT, on account
of its dimness and close proximity to the southern nuclear source. No
noteworthy $H{\alpha}$ features are associated with this feature
(Fig.\,8), excepting that B2 lies along the general disk of
$H{\alpha}$ that follows the galactic disk. More interesting is the
comparison with the CO emission (Fig.7) where it is seen that B2 lies
coincident with the CO feature corresponding to the south-western end
of the NGC4676B bar (also both B2 and the corresponding CO feature
show some extension to the northwest). A less bright X-ray feature
(not formally detected within the source-searching analysis) is seen
on the opposite side of the southern nucleus, at the other end of the
bar, and coincident with a smaller knot of CO emission. Gas dynamical
models of barred galaxies (\eg\ Engelmaier \& Gerhard 1997) show
strong gas accumulation at the ends of bars, due to corotation of the
bar structure with the disk, leading to enhanced star formation. The
CO emission and the general H$\alpha$ structure here seem to support
this, and source B2 and the X-ray feature on the other side of the bar
are very likely due to this bar-enhanced star-formation. Similar
behaviour has been seen in a number of other barred galaxies (\eg\
NGC4303; Tsch\"{o}ke, Hensler \& Junkes 2000).

Two sources are also seen in the tidal tails; N in the northern tail,
and S at the eastern edge of the southern. Interestingly, both sources
sit at similar distances along the tails from their respective
`heads'. No other significant X-ray features are seen associated with
the tidal tails. An enhancement in the ROSAT PSPC contours is seen at
the position of N (RP98), and, though nothing significant is seen at
the position of S, one would not expect to see anything significant,
given the brightness of S. The ROSAT HRI image (Fig.~1) shows an
interesting, though very low significance, plume to the north of
NGC4676A. None of the features within the plume however, coincide with
the position of N, and, given the ACIS-S count-rate of N, we would be
very fortunate to detect any emission in the HRI. Nothing is seen in
the HRI at the position of S.

N lies well within the northern tail, about 1.5\arcs\ east of the
centre of the tail. Though many $H{\alpha}$ features are seen within
both tails (Hibbard \& van Gorkom 1996), no bright knots lie
particularly close to N (the nearest is relatively dim, and lies some
1.5\arcs\ away). Interestingly, a knot in the HST ACS image is seen
poking eastwards out of the main central shaft of the tail at the
position of N. Furthermore there appears to be a neutral hydrogen \hi\
enhancement at the position of N (Yun \& Hibbard 2001).
S lies on the very eastern edge of the southern tidal tail. No
significant optical feature, either within the tail or more isolated,
appears coincident in the HST ACS image, or in coarser Digital Sky
Survey images. No $H{\alpha}$ features appear coincident either.

Finally, a bright star lies close to Source~W (visible at the very
right-hand edge of Fig.~2), but is too far offset to be
associated. Instead, a very faint optical counterpart can be seen at
the exact position of source~W in the HST ACS image. The ROSAT PSPC
contours are seen to stretch over to encompass this source (RP98), and
a feature is seen at this position in the ROSAT HRI data
(Fig.~1). This source is assumed not to be associated with the Mice,
and is likely a background AGN. 

Indeed, one can use the the logN$-$logS relation of Giacconi \etal\
(2001) to estimate the expected number of sources not physically
associated with the Mice. At most, only one background source is
expected over the area covered by Figs.\,2 \& 3
($\sim3.2$\,sq.\,\arcm) at the detection limit seen here
($5\times10^{39}$\,erg s$^{-1}$ at 88\,Mpc). It may also be the case
that Source S has nothing to do with the Mice, given that it lies some
distance from the southern tail, but for source N, lying as it does,
directly on the thin northern tail, the case is more concrete.

As a last point, it is worth mentioning that Chandra is able to throw
some light upon some of the curious X-ray features previously seen by
ROSAT, but not associated with the Mice. The strange extension to the
south-west (away from the tidal tails) seen in the PSPC image (RP98)
lies towards the direction of two dim Chandra X-ray sources; a
$\approx40$\,ct source at $\alpha=12^{\rm h}46^{\rm m}8.30^{\rm s}$,
$\delta=+30^{\circ}41\arcm 55.5\arcs $, and a $\approx20$\,ct source at
$\alpha=12^{\rm h}46^{\rm m}6.06^{\rm s}$, $\delta=+30^{\circ}42\arcm
24.2\arcs $. A final interesting source is the very dim
($\approx8$\,ct) source at $\alpha=12^{\rm h}46^{\rm m}10.42^{\rm s}$
$\delta=+30^{\circ}46\arcm 39.3\arcs $, lying directly along the line
of the northern tail, but approximately 1 arcminute beyond the optical
end of the tail.

Though only a small number of sources are evident within the Mice, an
approximate comparison of the source X-ray luminosity function (XLF)
with the XLFs of other galaxies can be made. This has recently been
performed for the Mice's more evolved and nearby counterpart, the
Antennae (Zezas \& Fabbiano 2002). At the distance of the Mice, one is
only able to detect sources with $L_{X}$\gtsim5$\times10^{39}$\,erg
s$^{-1}$. This in itself is important; all five sources detected
within the Mice are ultraluminous. This high tail to the XLF is in
sharp contrast to the Galaxy, and to a vast majority of nearby normal
and starburst galaxies (\eg\ Read \& Pietsch 2001, Zezas \& Fabbiano
2002). It does however, appear broadly similar to the Antennae. The
eight sources in the Antennae with $L_{X}>5\times10^{39}$\,erg
s$^{-1}$ may be significantly greater than the five in the Mice,
though there are a number of notes and caveats. Zezas \& Fabbiano
(2002) use a distance to the Antennae calculated using \ho\ = 50\,km
s$^{-1}$ Mpc$^{-1}$. Using \ho\ = 75\,km s$^{-1}$ Mpc$^{-1}$, the
eight high-$L_{X}$ Antennae sources would reduce to around five. Using
an Antennae distance of 25\,Mpc, as in RP98, the number would be
6$-$7. Furthermore, the Antennae is slightly more massive (in terms of
$L_{B}$; RP98), and has a correspondingly higher SFR (in terms of
$L_{FIR}$; RP98) by similar factors of $\approx1.3$.  Activity
indicators, such as far-infrared dust colour temperature
$S_{60}/S_{100}$, and (naturally, from above) $L_{FIR}/L_{B}$ are
almost identical for the two systems. Scaling the XLFs by mass (or
SFR) would bring the two systems into alignment. Note that the usage
of the optical luminosity $L_{B}$ as a measure of mass is generally
not ideal, as the blue luminosity to mass ratio is quite sensitive to
the age of a stellar population. Here however, as we are dealing with
two very similar systems, the comparison is quite justified (the usage
of various mass and activity indicators for galaxies is discussed in
Read \& Ponman 2001). Though it would seem therefore, that in terms of
the high end of the XLF, the Mice and the Antennae may appear rather
similar, a final caveat to consider is that it is likely, even with
the excellent spatial resolution of Chandra, that source confusion is
prevalent in the more distant Mice. Such confusion can only strengthen
the tail of the XLF, and the likely true situation is that there are
fewer than five high-$L_{X}$ sources in the Mice. It is difficult to
be conclusive, but it appears that there are significant number of
high-$L_{X}$ sources in the Mice, a number probably less than is seen
in the more evolved Antennae. If the Mice is to evolve into a system
like the Antennae, then it is expected that the number of high-$L_{X}$
sources in the Mice should increase slightly with time.

\subsection{Diffuse emission}

There exists much evidence that starburst-driven diffuse emission
features are seen around both galaxies within the Mice. As discussed
with regard to the point sources, there is very little doubt that a
good deal of star-formation is taking place at the two nuclei. Around
the two nuclei, one sees extended, clumpy, structured emission that is
seen to be soft and well-fitted with a thermal (0.5\,keV) spectral
model. Furthermore, emission appears to be significantly more extended
along the minor axis of each galaxy (this is more evident in the
northern galaxy, due likely to NGC4676A lying almost exactly
edge-on). Furthermore, in the northern galaxy, the H$\alpha$ emission
is also seen to be significantly extended along the minor axis
(especially to the west; Fig.\,8). All this is very reminiscent of
these features being starburst-driven galactic winds, as seen in
famous nearby starburst galaxies such as M82 and NGC253 (\eg\
Strickland \etal\ 2002; Bravo-Guerrero, Read \& Stevens 2002). Such
nearby systems however, are rather isolated, and classic bipolar winds
in even one member of a strong, rapidly-evolving interacting pair such
as the Mice, is something that has not been seen before. It is
believed from ROSAT (RP98) that starburst- and star formation-driven
diffuse emission is prevalent in interacting and merging galaxies, but
this was only seen in mid-stage (\eg\ the Antennae) and ultraluminous
(nuclear contact) mergers, not at early (\eg\ Mice) stages. In these
post-Mice cases, the systems are evolved to such a degree, that any
classic starburst winds (were they to have existed), would have been
distorted out of recognition, in agreement with observed peculiar
morphologies (RP98). More recently, higher-resolution Chandra
observations of the (post-Mice) Antennae show a great deal of hot
diffuse gas, but it has become all pervasive, extending further than
the stellar bodies of the galaxies (Fabbiano \etal\ 2002). It is
believed we have now seen, in the Mice, the onset of intense star
formation, starburst activity and starburst-driven diffuse gaseous
outflows in a classic full-blown disk-disk merger.

Though one cannot probe too far, due to lack of counts, there even
appears to be some temperature structure in the NGC4676A wind. The
eastern wind appears to be brighter and spectrally harder (greener,
rather than red; see Fig.\,2b) to the north than to the south. To the
west of NGC4676A, the situation is reversed, and all this may well be
an indication that the rapid evolution and slewing of the galaxies is
already beginning to effect even these very nascent starburst
winds. The existence of a classic starburst wind in a disk-disk
merger may be a very short-lived affair. 

One can infer mean physical properties of the hot gas around the
northern and southern galaxies once some assumptions have been made
regarding the geometry of the diffuse emission. The gas around each
galaxy is assumed to be contained in a spherical bubble of radius $r$,
taken to be the average radius of the lowest contour level seen in
Fig.2a, i.e. $\approx$7.5\arcs\ (for A) and $\approx$6.0\arcs\ (for
B). Using these volumes, the fitted emission measure $\eta n^{2}_{e}
V$ (where $\eta$ is the `filling factor' - the fraction of the total
volume $V$ which is occupied by the emitting gas) can be used to infer
the mean electron density $n_{e}$, and hence, assuming a plasma
composition of hydrogen ions and electrons, the total mass
$M_{\mbox{\small gas}}$ and thermal energy of the gas $E_{\mbox{\small
th}}$. Approximate values of the cooling time $t_{\mbox{\small cool}}$
of the hot gas, and also the mass cooling rate $\dot{M}_{\mbox{\small
cool}}$ and adiabatic expansion timescale $t_{\mbox{\small exp}}$, can
also be calculated. The resulting gas parameters for A and B are
listed in Table\,\ref{table_gas}.

\begin{table*}
\begin{center}
\begin{tabular}{ccccccccc}    \hline
Diff. & $kT$ & r & $n_{e}$ & $M_{\mbox{\small gas}}$ &
$E_{\mbox{\small th}}$ & $t_{\mbox{\small cool}}$ &
$\dot{M}_{\mbox{\small cool}}$ & $t_{\mbox{\small exp}}$ \\
Src.  &(keV) &(kpc)&(cm$^{-3}$)& ($M_{\odot}$)           & (erg)                  & 
(Myr)                    & ($M_{\odot}$ yr$^{-1}$)        &  (Myr)                   \\
      &      &     &($\times1/\sqrt{\eta}$) & ($\times\sqrt{\eta}$) &  ($\times\sqrt{\eta}$)   & 
($\times\sqrt{\eta}$)    &                                & \\ \hline
A & 0.50 & 3.2 & 0.019 & 6.5$\times10^{7}$ & 1.8$\times10^{56}$ & 480 & 0.13 & 8.8 \\
B & 0.46 & 2.6 & 0.025 & 4.3$\times10^{7}$ & 1.2$\times10^{56}$ & 380 & 0.11 & 7.3 \\
\hline
\end{tabular}
\caption{Values of physical parameters for the diffuse gas associated
with NGC4676A and NGC4676B. The values quoted are for a bubble model
(see text), and $\eta$ is the filling factor of the gas.}
\label{table_gas}
\end{center}
\end{table*}

Comparing the diffuse gas parameters in Table~3 with those for
isolated normal and starburst galaxies (Read, Ponman \& Strickland
1997), and for merging galaxies (RP98), one can say that the diffuse
outflows in NGC4676A and NGC4676B appear to be small, and probably
young. Extents of 9$-$14\,kpc are seen for the classic winds of M82
and NGC253, and the outflowing, turbulent ISM of the Antennae is seen
out to $\approx$8\,kpc (Fabbiano \etal\ 2002). Perhaps 5$-$10 times as
much gas mass is contained within the M82 or NGC253 diffuse features
than in either of the Mice, and the diffuse gas mass of the Antennae
may be up to 20 times that of the Mice. The outflows in the Mice
appear, in terms of their extent and mass, more like those of small
wind/large corona systems, such as NGC891 and NGC4631. 

In terms of temperature, it is believed that the hot ISM of starburst
galaxies has a a multi-temperature structure, and as such, given the
number of diffuse counts from the Mice, little conclusive can be
said. However, the single temperatures obtained from the spectral
fitting of the diffuse spectra ($\approx$0.50\,keV) is wholly
consistent with the range obtained for other starburst and merging
galaxies.

\subsection{X-ray emission from the Mice}

It has been seen that the X-ray emission from the Mice is made up of
point source and diffuse emission. In calculating a total $L_{X}$ for
the Mice, source W has been omitted, and the remaining 5 sources and
the two diffuse emission features have been summed together. Total
intrinsic (and emitted) 0.3$-$10\,keV X-ray luminosities (in units of
10$^{40}$\,erg s$^{-1}$) are as follows: Sources $-$ 5.56 (5.28),
Diffuse emission $-$ 1.88 (1.75), Total $-$ 7.43 (7.03), leading to a
fraction of the total emission that is diffuse of 25\%. As such it is
rather brighter, both fractionally and in absolute terms, than most
normal and starburst galaxies.

It is very instructive to compare these values with those of the
Antennae (Fabbiano \etal\ 2001), where approximately half of the total
(0.1$-$10\,keV) X-ray luminosity (2.3$\times10^{41}$\,erg s$^{-1}$) is
attributable to point sources, and half is due to the extended diffuse
thermal component. Given the uncertainties in the assumed distances to
both galaxies, and noting the slight difference in galaxy masses, it
appears that, whereas the X-ray point source populations in the two
systems may be rather similar (at least at the high-$L_{X}$ end), the
extended diffuse emission component is far fainter and less extended
in the Mice than it is in the more evolved Antennae. This indicates
that, if both systems are at different evolutionary stages of an
encounter involving similar gaseous disks, then at the Mice stage, the
ULX populations have evolved more rapidly than the diffuse gas
structures, whereas at the Antennae stage, the diffuse gas structures
have caught up. This is not surprising given that the ULXs are likely
due to the evolved components of single massive stars (\eg\ Roberts
\etal\ 2002), and thus can be created over short (10$^{6}$\,yr)
timescales, whereas the diffuse components are mainly due to galactic
winds, galactic fountains and chimneys, and the hot phases of the ISM,
requiring the agglomeration of several, perhaps thousands, of
supernova remnants, thus requiring a far longer (few
$\times10^{7}$\,yr) timescale.

The Mice and the Antennae appear to be at different stages of very
similar evolutionary tracks (involving two similarly-sized gas-rich
spiral disks), the only difference between the systems being (apart
from their merger epoch) that the Antennae appears slightly more
massive. It is therefore attractive to predict that that the Mice will
evolve into a system very like the Antennae within the next
$10^{8}$\,years or so. Though the high-$L_{X}$ end of the point source
populations of the two systems are already very similar, the diffuse
gas component of the Mice's emission requires a few
$\times10^{7}$\,years worth of starburst-injected mass and energy,
before it can compare with the diffuse emission seen in the
Antennae. At present, the outflows in the Mice, in terms of their size
and mass, are not yet especially significant, resembling structures
observed in nearby systems with small winds and/or large coronae.

\section{Conclusions}

Presented here are high spatial and spectral resolution Chandra ACIS-S
X-ray observations of the famous interacting galaxy pair, the Mice, a
system very similar to the Antennae galaxies (albeit at a much larger
distance), in consisting of the merger of two equally-sized spiral
galaxies. The Mice are at an earlier interaction stage however, the
galaxy disks not yet having begun to merge. Previously unpublished
ROSAT HRI data of the system are also presented. The primary results
can be summarized as follows:

\begin{itemize}

\item Of great interest is the discovery of what appears to be
starburst-driven galactic winds outflowing along the minor axes of
both galaxies (particularly the northern). Spectrally soft X-ray
emission extends beyond the northern edge-on galactic disk, into
regions occupied by plumes of H$\alpha$ emission. Fitting of the
diffuse emission spectra indicates temperatures wholly consistent with
well-known nearby galactic winds. That such classic winds could exist
in the Mice is perhaps surprising, as one would expect the rapid
evolution and violence of the environment to distort the wind
structures very quickly. The phenomenon may not last long however $-$
in terms of extent and mass, the winds are small, low-luminosity,
perhaps newly-formed and already show complex spatial and spectral
structure, indicating that they could be being disrupted by the
encounter. These low-temperature, diffuse X-ray features, extending
out of the galactic disks appear to be the very beginnings of
starburst-driven hot gaseous outflows in a full-blown disk-disk
merger.

\item In addition, five bright ($L_{X}$ (0.3$-$10\,keV) \gtsim
5$\times10^{39}$\,erg s$^{-1}$) point sources are formally detected,
associated with the Mice system. The sources detected at the nuclei of
the two galaxies are very bright, though perhaps not single,
individual sources, and appear quite similar. Emission from the
northern nucleus is likely absorbed by the intervening edge-on disk of
the northern galaxy. Both nuclear X-ray sources, given the spatial
structure of the nuclear and surrounding diffuse X-ray emission, and
the large amounts of associated molecular gas and H$\alpha$ emission,
are likely starburst regions. A source is detected coincident with a
molecular gas complex at the tip of one edge of the southern galaxy's
bar. A smaller X-ray enhancement (together with molecular emission) is
observed at the opposite end. Two further spectrally soft sources are
detected within the two tidal tails.

\item The source X-ray luminosity function appears in sharp contrast
to that of other normal and starburst galaxies. It is similar however
to that of the more evolved Antennae, though the number of
high-$L_{X}$ sources in the Mice is probably less.

\item Whilst the point source XLFs of the Mice and Antennae appear
similar, far less diffuse emission is detected in the Mice. This
indicates that the high-luminosity source populations in these systems
evolve more rapidly than the diffuse gas structures. For the Mice to
evolve, within the next $10^{8}$\,years, into a system like the
Antennae, much of the starburst-injected energy and mass needs to feed
the diffuse gas component.

\item There is no evidence for any variability between the Chandra
data and the ROSAT HRI data.

\end{itemize}

\section*{Acknowledgements} 

AMR acknowledges the support of PPARC funding, and thanks the referee
for useful comments which have improved the paper. AMR also thanks
John Hibbard for making his H$\alpha$ data available, and Trevor
Ponman for carefully reading the manuscript. Optical images are based
on photographic data obtained with the UK Schmidt Telescope, operated
by the Royal Observatory Edinburgh, and funded by the UK Science and
Engineering Research Council, until June 1988, and thereafter by the
Anglo-Australian Observatory.  Original plate material is copyright
(c) the Royal Observatory Edinburgh and the Anglo-Australian
Observatory. The plates were processed into the present compressed
digital form with their permission. The Digitized Sky Survey was
produced at the Space Telescope Science Institute under US Government
grant NAG W-2166.


\begin{thebibliography}{99}

\bibitem{} Barnes J.E., Hernquist L., 1991, ApJ, 370, L65
\bibitem{} Barnes J.E., Hernquist L., 1996, ApJ, 471, 115
\bibitem{} Bravo-Guerrero J., Read A.M., Stevens I.R., submitted to MNRAS
\bibitem{} Engelmaier P., Gerhard O., 1997, MNRAS, 287, 57
\bibitem{} Fabbiano, G., Zezas, A. Murray, S.S., 2001, ApJ, 554, 1035
\bibitem{} Giacconi R., \etal, 2001, ApJ, 551, 624 
\bibitem{} Heckman T.M., 1993, in Schlegel E.M., Petre R., eds, AIP
	Conf. Proc. 313, The Soft X-ray Cosmos. AIP Press, Woodbury, NY, p.139
\bibitem{} Hibbard J.E., van Gorkom J.H., 1996, AJ, 111, 655
\bibitem{} Mihos J.C., Bothun G.D., Richstone D.O., 1993, ApJ, 418, 82
\bibitem{} Read A.M., Pietsch W., 2001, A\&A, 373, 473 
\bibitem{} Read A.M., Ponman T.J., D.K., Strickland D.K., 1997, MNRAS, 286, 626
\bibitem{} Read A.M., Ponman T.J., 1998, MNRAS, 297, 143 (RP98)
\bibitem{} Read A.M., Ponman T.J., 2001, MNRAS, 328, 127 
\bibitem{} Roberts T.P., Warwick R.S., Ward M.J., Murray S.S., 2002, MNRAS, 337, 677  
\bibitem{} Schombert J.M., Wallin J.F., Struck-Marcell C., 1990, AJ, 99, 497
\bibitem{} Schweizer F., 1989, Nat, 338, 119
\bibitem{} Smith B.J., Higdon J.L., 1994, AJ, 108, 837
\bibitem{} Sotnikova N.Y., Reshetnikov V.P., 1998, Astronomy Letters, 24, 73
\bibitem{} Stockton A., 1974, AJ, 187, 219
\bibitem{} Strickland D.K., Heckman T.M., Weaver K.A., Hoopes C.G.,
	Dahlem M., 2002, ApJ, 568, 689
\bibitem{} Toomre A., 1977, in Evolution of Galaxies and Stellar Populations, Ed. 
        B.M.Tinsley, R.B.Larson, p.401, (New Haven: Yale University Observatory) 
\bibitem{} Toomre A., Toomre J., 1972, ApJ, 178, 623 
\bibitem{} Tsch\"{o}ke D., Hensler G., Junkes N., 2000, A\&A, 360, 447
\bibitem{} Vorontsov-Vel'yaminov B.A., 1957, Astron. Tsirk., 178, 19
\bibitem{} Young J.S., Scoville N.Z., 1991, ARA\&A, 29, 581
\bibitem{} Yun M.S., Hibbard J.E., 2001, ApJ, 550, 104
\bibitem{} Zezas A., Fabbiano G., 2002, 2002, ApJ, 577, 726
\bibitem{} Zezas A., Fabbiano G., Rots A.H., Murray S.S., 2002, ApJ, 577, 710

\end{thebibliography}
\end{document}